\documentclass[twocolumn,prl,floats,showpacs,superscriptaddress]{revtex4}
\usepackage{graphicx,epsfig}
\usepackage{times}
\usepackage{graphics,dcolumn,bm,float}
\usepackage{amssymb,amsmath,rotate,color}

\begin{document}
\unitlength 1 cm
\newcommand{\be}{\begin{equation}}
\newcommand{\ee}{\end{equation}}
\newcommand{\bearr}{\begin{eqnarray}}
\newcommand{\eearr}{\end{eqnarray}}
\newcommand{\nn}{\nonumber}
\newcommand{\vk}{\vec k}
\newcommand{\vp}{\vec p}
\newcommand{\vq}{\vec q}
\newcommand{\vkp}{\vec {k'}}
\newcommand{\vpp}{\vec {p'}}
\newcommand{\vqp}{\vec {q'}}
\newcommand{\bk}{{\bf k}}
\newcommand{\bp}{{\bf p}}
\newcommand{\bq}{{\bf q}}
\newcommand{\br}{{\bf r}}
\newcommand{\bR}{{\bf R}}
\newcommand{\up}{\uparrow}
\newcommand{\down}{\downarrow}
\newcommand{\fns}{\footnotesize}
\newcommand{\ns}{\normalsize}
\newcommand{\cdag}{c^{\dagger}}

\title{The effect of vacancy-induced magnetism on electronic transport in armchair carbon nanotubes}
\author{R. Farghadan}
\affiliation{Department of Physics, Tarbiat Modares University, Tehran, Iran }
\author{A. Saffarzadeh} \email{a-saffar@tpnu.ac.ir}
\affiliation{Department of Physics, Payame Noor University,
Nejatollahi Street, 159995-7613 Tehran, Iran}
\affiliation{Computational Physical Sciences Laboratory,
Department of Nano-Science, Institute for Research in Fundamental
Sciences (IPM), P.O. Box 19395-5531, Tehran, Iran}
\date{\today}

\begin{abstract}
The influence of local magnetic moment formation around three
kinds of vacancies on the electron conduction through metallic
single-wall carbon nanotubes is studied by use of the Landauer
formalism within the coherent regime. The method is based on the
single-band tight-binding Hamiltonian, a surface Green's function
calculation, and the mean-field Hubbard model. The numerical
results show that the electronic transport is spin-polarized due
to the localized magnetic moments and it is strongly dependent on
the geometry of the vacancies. For all kinds of vacancies, by
including the effects of local magnetic moments, the electron
scattering increases with respect to the nonmagnetic vacancies
case and hence, the current-voltage characteristic of the system
changes. In addition, a high value for the electron-spin
polarization can be obtained by applying a suitable gate voltage.

\end{abstract}

\maketitle

\section{Introduction}
The electrical transport properties of single-wall carbon
nanotubes (CNTs) and other carbon-based materials have attracted
much attention due to their unusual properties and great potential
for technological applications
\cite{Nardelli,Rochefort,Berger,Suarez}. Among these features, the
ballistic electron conduction and the long range spin coherent
transport for perfect and defective single-wall CNTs have been
investigated theoretically and experimentally
\cite{Ando,Frank,Tsukagoshi}. These important properties of spin
polarized electrons in single-wall CNTs has motivated their use in
the emerging field of spin electronics \cite {Son} which aims to
effectively control and manipulate the spin degrees of freedom in
the electronic devices \cite{Zutic}.

In order to achieve spintronic devices based on single-wall CNTs,
it is important to understand the magnetic effect of vacancies and
impurities on the electron conduction \cite{Son,Lehtinen,Sancho}.
because, the electronic properties of carbon-based materials
strongly depend on their topological structure \cite{Igami}.
Therefore the electronic structure of the CNTs can differ due to
the topological defects or the addition of different compounds. On
the other hand, in the transport processes, the ballistic
conductance depends on the number of conducting channels at the
Fermi energy  \cite{Fisher}. Consequently, the appearance of the
vacancy defects in a structure can change the electronic and
transport properties of the system \cite{Ando}. Also, the
conductance of an imperfect system is lowered due to the
reflection of electron waves from the defects \cite{ Chico}. In
addition, the localized states near the vacancy are magnetic and
change the net magnetic moments in the carbon-based
nanostructures. The honeycomb lattice of graphene sheet is formed
by two sublattices $A$ and $B$ (bipartite lattice). For a
bipartite lattice with different numbers $N_A$ and $N_B$ of sites
and the Hubbard repulsive parameter, the total spin, $S$, of the
ground state of the system, which is mainly localized near the
vacancy, is $2S={N_A}-{N_B}$ \cite{Lieb}. This important feature
can block and change the spin transport especially near the Fermi
energy.

In this regard, the electronic and magnetic properties of
vacancies in single-wall CNTs \cite{Orellana} and graphene
nanoribbons \cite{Kumazaki} have been investigated. In addition,
the electron-spin polarization has been observed in the CNTs when
doped with magnetic adatoms or molecules \cite{Yang}. Moreover,
the localized states of the impurity can change the spin-polarized
conduction in the presence of a gate voltage or applied bias
\cite{Son,Fedorov}. Recently, the effect of vacancy on the
conductance of single-wall CNTs \cite{Choi,zhang} and the
spin-dependent transport properties in ferromagnetically contacted
single-wall CNTs have been investigated \cite{Kim,Jensen,Fedorov},
but the magnetic behavior of vacancy defect as regards the
transmission of single-wall CNTs has not been considered.

The purpose of this work is to study the effect of magnetic
vacancies on the spin-polarized transport through armchair CNT
junctions and manipulate this polarization by means of gate
voltages. We simulate ideal vacancies which are made by removing
carbon atoms from lattice sites without including the deformation
of the tube wall around the vacancies. We consider three typical
vacancy types. For the first type, a single carbon site (A or B)
is removed. For the second one, two same sites (two A or B sites),
and in the last type, four A sites in one carbon ring are removed,
as shown in Figs. 1(a) and 1(b), respectively.
\begin{figure}
\centerline{\includegraphics[width=0.95\linewidth]{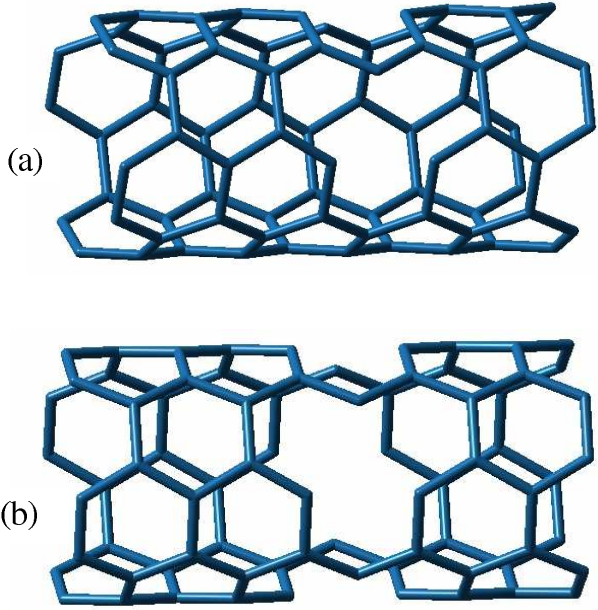}}
\caption{Five unit cells of a (4,4) single-wall CNT (consist of 80
atoms) as the channel, in the presence of (a) two and (b) four
vacancies on a ring of carbon atoms.}
\end{figure}

Using the single-band tight-binding approximation and the
mean-field Hubbard model \cite {Fujita}, the electronic structure
and the localized magnetic moments around the above-mentioned
vacancies are calculated. Also, using the non-equilibrium Green's
function technique and the Landauer-B\"{u}ttiker theory
\cite{Datta}, the spin-polarized transport in defective
single-wall CNTs is investigated.

\section{Model and formalism}
We consider a system that consists of a defective CNT with finite
length, sandwiched between two perfect semi-infinite CNTs as
electrodes. All CNTs have same structure, (n,n). To calculate the
electron transmission properties of this defective CNT which acts
as a channel, the electronic structure of this system should be
resolved in detail. Hence, we decompose the total Hamiltonian of
the system as \cite{Saffarzadeh}
\begin{equation}
\hat{H}=\hat{H}_{L}+\hat{H}_{C}+\hat{H}_{R}+\hat{H}_{T}\  ,
\end{equation}
where $\hat{H}_L$ and $\hat{H}_R$ are the Hamiltonians of the left
($L$) and right ($R$) electrodes, $\hat{H}_C$ describes the
channel Hamiltonian and contains the magnetic properties of the
vacancies, and $\hat{H}_T$ is the coupling between the electrodes
and the central part (defective CNT). The total Hamiltonian is
described within the single-band tight-binding approximation.
Therefore, the Hamiltonian of the electrodes can be written as
\cite{Esfarjani}
\begin{equation}
\hat{H}_{\alpha}=\sum_{<i_\alpha,
j_\alpha>,\sigma}(\epsilon_{\alpha}\delta_{i_\alpha,j_\alpha}-t_{i_\alpha,j_\alpha})
\,\hat{c}_{i_\alpha,\sigma}^\dag\hat{c}_{j_\alpha,\sigma}\ ,
\end{equation}
where $\hat{c}_{i_\alpha,\sigma}^\dag$
($\hat{c}_{i_\alpha,\sigma}$) creates (destroys) an electron at
site $i$ in electrode $\alpha(=R,L)$ and the hopping elements
$t_{i_\alpha,j_\alpha}$ are equal to $t$ for nearest neighbors and
zero otherwise. Here, $\epsilon_{\alpha}$ is the on-site energy of
the electrodes and will be set to zero. The coupling Hamiltonian
is described as
\begin{equation}
\hat{H}_{T}=-\sum_{\alpha=\{L,R\}}\sum_{i_\alpha,
j_C,\sigma}t_{i_\alpha,j_C}(\hat{c}_{i_\alpha,\sigma}^\dag\hat{d}_{j_C,\sigma}+\textrm{H.c.})\
\ ,
\end{equation}

The parameters $t_{i_\alpha,j_C}$ for hopping between the
electrodes and the channel are taken to be $t$. In order to obtain
the magnetic moment of each atom around a vacancy, we use the
Hubbard model within the unrestricted Hartree-Fock approximation
\cite{Fujita}. Accordingly, the Hamiltonian of the central part
within the mean-field approximation of the Hubbard model can be
written as \cite{Fujita, Rossier}
\begin{eqnarray}\label{4}
\hat{H}_{C}=\sum_{i\sigma}\epsilon_i\hat{d}_{i\sigma}^\dag\hat{d}_{i\sigma}-t\sum_{i,j,\sigma}\hat{d}_{i\sigma}^\dag\hat{d}_{j\sigma}
+U\sum_i\{\langle \hat{n}_{i\uparrow}\rangle
\hat{n}_{i\downarrow}\nonumber\\ +\langle
\hat{n}_{i\downarrow}\rangle \hat{n}_{i\uparrow}-\langle
\hat{n}_{i\uparrow}\rangle\langle\underline{}
\hat{n}_{i\downarrow}\rangle\}\ ,
\end{eqnarray}
where $\epsilon_i$ is the on-site energy and will be set to the
gate potential in the channel region, the second term corresponds
to the single $\pi$-orbital tight-binding Hamiltonian, while the
third term accounts for the on-site Coulomb interaction $U$. In
this expression, $\hat{d}_{i\sigma}^\dag(\hat{d}_{i\sigma})$
creates (annihilates) an electron, and $\langle
\hat{n}_{i\sigma}\rangle$ is the expectation value of the number
operator for an electron with spin $\sigma$ at the $i$th site.

The Green's function of the system can be written as
\begin{equation}
\hat{G}_{C}(\varepsilon)=[\varepsilon\hat{I}-\hat{H}_{C}-\hat\Sigma_{L}(\varepsilon)-\hat\Sigma_{R}(\varepsilon)]^{-1}\
,
\end{equation}
where
$\hat{\Sigma}_{\alpha}(\varepsilon)=\hat{H}_{C\alpha}{\hat{g}}_{\alpha}(\varepsilon)\hat
{H}_{C\alpha}^\dag$ is the self-energy matrix which contains the
influence of the electronic structure of the semi-infinite
electrodes through the lead's surface Green's function $g_\alpha$
\cite{Lopez, Lee}. Correspondingly, $\hat{H}_{C\alpha}$ defines
the matrix of coupling between the surface atomic orbitals of the
lead $\alpha$ and the channel. We should note that, $\hat{G}_C$ is
a spin-dependent matrix of size 2$N$, with $N$ being the number of
atoms of the central region. Therefore, due to our mean-field
decoupling scheme, we can decouple the electronic transport into
spin-up and spin-down currents.

Using the iterative method for the calculation of the transfer
matrices based on the principal layers concept, the surface
Green's function of the left and right leads can be obtained as
\cite{Lopez}:
\begin{eqnarray}
{\hat{g}}_{L}(\varepsilon)=[\varepsilon\hat{I}-\hat{H}_{00}^{L}-\hat{H}_{01}^{L}\,{\hat{\bar{T}}}_{L}]\ ,\nn\\
{\hat{g}}_{R}(\varepsilon)=[\varepsilon\hat{I}-\hat{H}_{00}^{R}-\hat{H}_{01}^{R}\,{\hat{T}}_{R}]\
,
\end{eqnarray}
where $\hat{H}_{00}^{\alpha}$ and $\hat{H}_{01}^{\alpha}$ are
matrices and correspond to an isolated principal layer and the
interaction between two nearest principle layers in lead $\alpha$,
respectively. Also, ${\hat{\bar{T}}}_{L}$ and ${\hat{T}}_{R}$ are
the transfer matrices and for the details of the calculations the
reader can refer to Ref \cite{Lopez}. From the expression of the
non-equilibrium Green's function, the spin-dependent local density
of states (LDOS) and the expectation value for the number operator
of electrons in the channel are given by
\begin{equation}
D_{i\sigma}(\varepsilon)=-\frac{1}{\pi}\,\mathrm{Im}\,(\langle{i\sigma}\textemdash\hat{G}_{C}(\varepsilon)\textemdash{i\sigma}\rangle)\
,\nn
\end{equation}
\begin{equation}\label{n}
\langle{\hat{n}}_{i\sigma}\rangle=\int_{-\infty}^{E_F}{D_{i\sigma}(\varepsilon)}d{\varepsilon}\
.
\end{equation}
Accordingly, the magnetic moment, $m_i$, at site $i$ of the
channel can be calculated using
\begin{equation}
m_{i}=\mu_B\,[\langle \hat{n}_{i\uparrow}\rangle-\langle
\hat{n}_{i\downarrow}\rangle]\ .
\end{equation}

In this study, we solve the mean field Hamiltonian
self-consistently by an iterative method \cite{Fujita}. In the
first step, we start from an anti-ferromagnetic configuration as
an initial condition and establish the Hamiltonian for the Hubbard
model [Eq. (\ref{4})]. In the second step, the effect of
electrodes on the channel is added via $\hat\Sigma_L$ and
$\hat\Sigma_R$, and then the Green's function of the channel,
$\hat G_C$, in the presence of the electrodes is calculated. In
the third step, the expectation values of the number operators
$\langle{\hat{n}}_{i\sigma}\rangle$ at each site and for both spin
alignments are calculated by using the Green's function. Finally
the new expectation values of the number operators are replaced in
Eq. (\ref{4}), and this process is repeated until the difference
between two successive iterations becomes less than $10^{-4}$.

The total Hamiltonian does not contain inelastic scattering,
because there is no spin-flip or the other sources of scattering
in the system.  In other words, the transmission probabilities of
majority and minority subbands are independent and the electronic
transport can be decoupled into two spin currents: one for spin-up
and the other for spin-down. Therefore, the spin-dependent
currents for a constant bias voltage are calculated by using the
Landauer-B\"{u}ttiker formalism \cite{Lehtinen,Saffarzadeh}:
\begin{equation}
I_\sigma(V_a)=\frac{e}{h}\int_{-\infty}^{\infty}
T_\sigma(\varepsilon)[f(\varepsilon-\mu_L)-f(\varepsilon-\mu_R)]d\varepsilon
\ ,
\end{equation}
where $f$ is the Fermi-Dirac distribution function,
$\mu_{L,R}=E_F\pm\frac{1}{2}eV_a$ are the chemical potentials of
the electrodes and
$T_\sigma(\varepsilon)=\mathrm{Tr}[\hat{\Gamma}_{L}
\hat{G}_{C,\sigma}\hat{\Gamma}_{R}\hat{G}_{C,\sigma}^{\dagger}]$
is the energy- and spin-dependent transmission function. Using
$\hat\Sigma_{\alpha}$, the coupling matrices $\hat\Gamma_\alpha$
can be expressed as
$\hat\Gamma_\alpha=-2\,\mathrm{Im}[\hat\Sigma_{\alpha}(\varepsilon)]$.

\section{results and discussion }
We study the LDOS, the transmission and the current-voltage
characteristic of the single-wall CNTs in the presence of
vacancies when the CNT in all regions is of (4,4) type. We expect
the magnetic behavior of the vacancy to affect the electron
conduction through the structure, since different scatterings
occur for the electronic waves with different spin densities.
Therefore, the transmission for majority electrons can be
different from that of the minority electrons at certain energies.

\begin{figure}
\centerline{\includegraphics[width=0.85\linewidth]{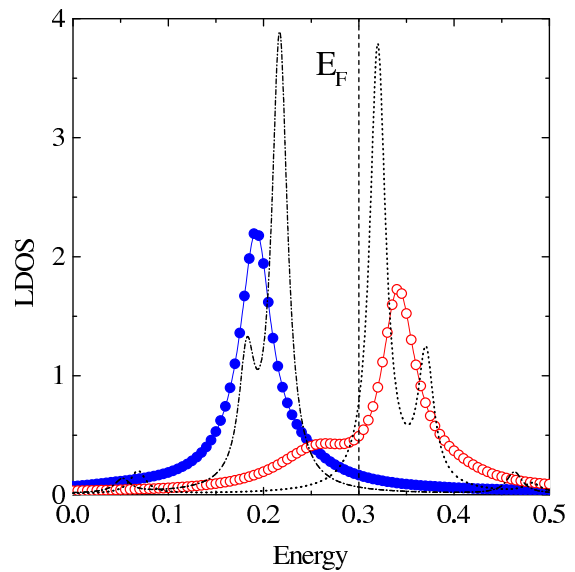}}
\caption{ LDOS per site around two vacancies for the
CNT/defective-CNT/CNT junction with $U=1.06\,t$, $V_G=0.0\,t$, and
$V_a=0.0\,t$. Full (open) circles are for majority (minority) spin
electron. The dotted (dashed-dotted) line is for majority
(minority) spin in the absence of CNT leads ( and the dashed line
indicates the Fermi energy). Note that, the majority and minority
electrons are determined by Eq. (\ref{n}).}
\end{figure}

\begin{figure}
\centerline{\includegraphics[width=0.85\linewidth]{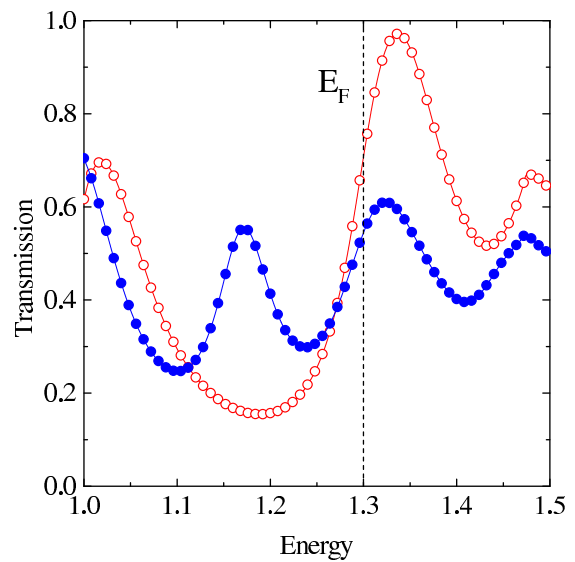}}
\caption{Transmission coefficients as a function of energy in the
presence of two vacancies (second type) with $V_G=1.0\,t$ and
$V_a=0.0\,t$. Full (open) circles are for majority (minority) spin
electrons.}
\end{figure}

In the cases of the first, second and third vacancy types, the
total magnetic moments for the finite CNT (channel) as shown in
Fig. 1, are 1, 2 and 4$\mu_B$ respectively, which are in good
agreement with the Lieb's theorem \cite{Lieb}. However, when the
two semi-infinite CNTs (electrodes) are attached to the channel,
their electronic properties are changed. This causes appreciable
variation in the value of magnetic moment of each atomic site near
the vacancies and affects the spin-dependent transport through the
channel. When the atoms from the same sublattice are removed, the
final spin configuration, after self-consistent calculations for
the atomic sites near the vacancies, is ferromagnetic. These
magnetic moments only obtain for the atomic sites near the
vacancies and for the other sites approximately vanish. Our
numerical results showed that in the presence of a single vacancy,
the magnetic moment vanishes approximately and an unpolarized
ground state is obtained for the value of $U$=1.06 $t$, due to the
effect of semi-infinite electrodes on the channel \cite{Rojas}.

In order to clarify this effect, we have plotted in Fig. 2 the
LDOS in the presence of two vacancies. In a defective CNT, the
electronic states on the carbon atoms near the vacancy strongly
depend on the presence or absence of CNT electrodes. In a
finite-length CNT, the electronic states on carbon atoms are
almost localized due to the absence of periodicity along the axis
of CNT. Therefore, sharp features are observed in the
spin-dependent LDOS for the states around the Fermi energy. When
the electrodes are attached to the defective CNT, the localization
of electronic states is suppressed and the sharp peaks are
broadened for both majority and minority electrons. Accordingly,
the difference between the two densities of states and hence the
spin polarizations are reduced. Note that in both cases the
condition $\langle \hat{n}_{i\uparrow}\rangle+\langle
\hat{n}_{i\downarrow\rangle}=1$ at each site is satisfied.

It should be pointed out that the Hubbard repulsive parameter
shifts the Fermi energy from zero to $E_F=0.3\,t$ as shown with a
vertical line in Fig. 2. In fact, when the two vacancies in the
channel are included in the calculations, we obtain $S$=0.55 for
the total spin value in the presence of CNT electrodes and the
maximum value of the magnetic moment reaches 0.16\,$\mu_B$. This
magnetic moment can affect the spin transport in the channel.

To see the effects of vacancy-induced magnetism on the electron
conduction, we have shown in Fig. 3 the transmission coefficients
as a function of energy for majority and minority electrons at
$V_G$=1.0\,$t$. The transmission spectrum of two spin subbands is
non-degenerate for all electronic states within the energy window
and there is a significant difference between the two spin
subbands. Since in the process of electronic transport, the
electrons with Fermi energy carry most of the current, we have
shown the transmission coefficients around the Fermi level. It is
clear that the Fermi energy shifts from $0.3\,t$ to $1.3\,t$ due
to the gate voltage.

We note that a gate voltage shifts the electronic states and
hence, the transmission for two spin subbands in the central
region can vary significantly. Therefore, a gate potential is able
to change the spin transport through such a defective CNT. In
order to see the effect of the gate voltage on the spin transport,
we have calculated the degree of spin polarization $(P)$ for
electrons traversing the channel which can be defined as
$P(E)=\frac{T_\uparrow(E)-T_\downarrow(E)}{T_\uparrow(E)+T_\downarrow(E)}$.
Fig. 4 shows this physical quantity in terms of the gate voltage
in the case of two vacancies. It is clear that the spin
polarization can reach values as high as 65\%. Also, we found that
the change in the position of the two vacancies with respect to
each other causes appreciable variations in the value of the
localized magnetic moment and hence, in the degree of
spin-polarization. Generally, on increasing the distance between
two vacancies the total value of the channel magnetic moment
decreases.

\begin{figure}
\centerline{\includegraphics[width=0.85\linewidth]{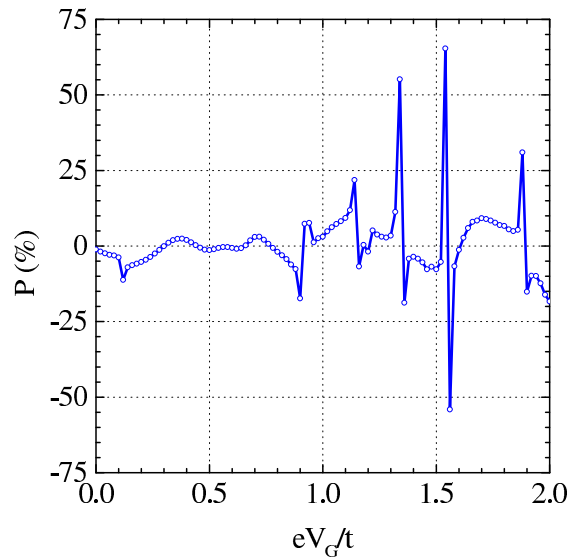}}
\caption{Degree of spin polarization as a function of gate voltage
for vacancies of the second type at the Fermi energy and
$V_a=0.0\,t$.}
\end{figure}

\begin{figure}
\centerline{\includegraphics[width=0.85\linewidth]{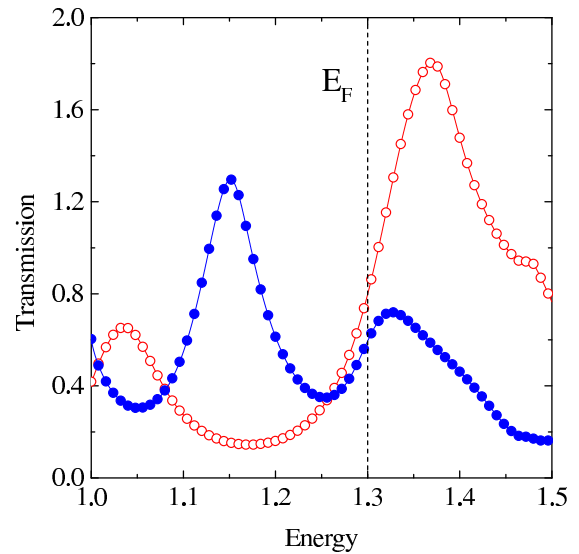}}
\caption{Transmission coefficients as a function of energy in the
presence of four vacancies (third type), for $V_G=1.0\,t$ and
$V_a=0.0\,t$. Full (open) circles are for majority (minority) spin
electrons.}
\end{figure}

Fig. 5. shows the transmission coefficients as a function of
energy for majority and minority electrons in the case of four
vacancies (third type). Only the energy window around the Fermi
level has been shown $(E_F=1.3\,t)$. As we know, in the present
structure where the electrodes are made of CNTs, there are more
paths for the electrons to pass from the electrodes to the
defective CNT. Therefore, in some energy ranges of this figure,
the spin-dependent transmission coefficients can be larger than
unity. The total value of the spin is $1.40$  and the maximum
value of the magnetic moment reaches $0.30\,\mu_B$. Also the
maximum value of the spin polarization at $V_G=1.0\,t$, reaches
80\% which indicates that the local magnetism induced by vacancies
can be useful in spintronic devices. This spin polarization of the
electrical conductivity can play an important role in the
current-voltage characteristics in the channel.

For this reason and to further understand the effect of
vacancy-induced magnetism, we have plotted in Fig. 6 the spin
currents in terms of the applied voltage. Since the electron
conduction depends on the electronic states lying between $\mu_L$
and $\mu_R$, on increasing the applied voltage, these states move
inside the energy window and the electronic currents for both
majority and minority spin electrons increase. From the
current-voltage characteristic, we see that the applied voltage
can change the difference between both spin-up and spin-down
currents significantly and therefore, we should adjust the value
of the gate and applied voltages to obtain maximum values for the
degree of polarization of the spin currents.

In this study, we only focused on the A-site defects because
according to Lieb's theorem, when the same sublattice atoms are
removed the total magnetic moment increases and therefore the
polarization in the spin currents becomes remarkable. Also, our
numerical results showed that if we remove two different sites
from one carbon ring, the local magnetic moments near the
vacancies are nonzero, while the total magnetic moment is zero and
hence, the transmission coefficients of two spin channels are
degenerate and practically unpolarized. In other cases, i.e., when
different numbers of A and B sites are removed, the total magnetic
moment is small and a little spin polarization can be observed
(not shown here).

\section{Conclusion}
In this work, the coherent spin-polarized transport through a
CNT/defective-CNT/CNT junction was investigated on the basis of
the non-equilibrium Green's function technique, and the mean-field
Hubbard model. We have shown that there are major differences in
the behavior of the electronic transport when different kinds of
vacancies are considered. The numerical results indicate that, the
point defects of which different types were considered-and hence
with different magnetic responses-could effectively alter the
electron conduction through the junction.
\begin{figure}
\centerline{\includegraphics[width=0.85\linewidth]{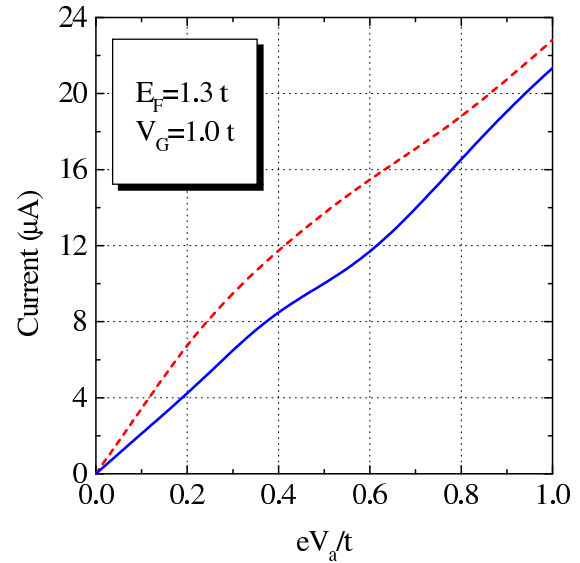}}
\caption{Current-voltage characteristics in the presence of four
vacancies. The solid (dashed) line is for majority (minority) spin
electrons.}
\end{figure}

In the defective CNTs, the difference between the transmission
coefficients of two spin subbands at some energy values is small.
Consequently, in such a case, the degree of spin polarization can
be manipulated by applying a gate voltage to the channel. In
addition, our results suggest that by choosing the position of the
vacancy in the channel properly to enhance the local
magnetization, one can increase the values of the spin-polarized
currents. These defects may be useful in the process of spin
injection into semiconducting devices for spintronic applications.

Throughout the study, we have ignored the effect of spin-flip
scattering, especially near the vacancies. This factor may affect
the spin-dependent transport, and hence, a further improved
approach is needed to obtain more accurate results.

\section*{Acknowledgement}
The authors would like to thank J. J. Palacios, K. Wakabayashi, M.
A. H. Vozmediano, and M. B. Nardelli for helpful discussions.

\end{document}